%% file: main.tex
\documentclass[11pt,a4paper]{article}

\usepackage[utf8]{inputenc}
\usepackage[english]{babel}
\usepackage{amsmath, amssymb, amsthm}
\usepackage{graphicx}

\usepackage{geometry}
\usepackage{cite}
\usepackage{booktabs}
\usepackage{caption}
\usepackage{subcaption} 
\usepackage{orcidlink}
\usepackage{algorithm}
\usepackage{algorithmicx}
\usepackage{algpseudocode} 

\usepackage{hyperref}

\title{\textbf{Are Enterprises Ready for Quantum-Safe Cybersecurity?}}

\author{
  Tran Duc Le$^{1}$ \\
  \texttt{let@uwstout.edu,\orcidlink{0000-0003-3735-0314}} \\
  \and
  Phuc Hao Do$^{2,*}$ \\  
  \texttt{do.hf@sut.ru,\orcidlink{0000-0003-0645-0021}} \\
  \and
  Truong Duy Dinh$^{3}$ \\
  \texttt{duydt@ptit.edu.vn,\orcidlink{0000-0002-9993-9792}} \\
  \and
  Van Dai Pham$^{4}$ \\
  \texttt{daipv11@fe.edu.vn,\orcidlink{0000-0003-1363-0784}} \\
}

\date{
    $^{1}$Mathematics, Statistics and Computer Science, University of Wisconsin–Stout, USA \\
    $^{2}$Department of Telecommunication Engineering, Bonch-Bruevich St. Petersburg State University of Telecommunications, Russia \\
    $^{3}$Posts and Telecommunications Institute of Technology, Vietnam \\
    $^{4}$Swinburne Vietnam, FPT University, Vietnam \\[2ex]
    $^{*}$Corresponding author: do.hf@sut.ru \\[2ex]
}

\begin{document}

\maketitle

\begin{abstract}
Quantum computing threatens to undermine classical cryptography by breaking widely deployed encryption and signature schemes. This paper examines enterprise readiness for quantum-safe cybersecurity through three perspectives: (i) the technologist view, assessing the maturity of post-quantum cryptography (PQC) and quantum key distribution (QKD); (ii) the enterprise (CISO/CIO) view, analyzing organizational awareness, risk management, and operational barriers; and (iii) the threat actor view, evaluating the evolving quantum threat and the urgency of migration. Using recent standards (e.g., NIST's 2024 PQC algorithms), industry surveys, and threat intelligence, we synthesize findings via a SWOT analysis to map strengths, weaknesses, opportunities, and threats. Results indicate uneven and generally insufficient preparedness: while PQC standards and niche QKD deployments signal technical progress, fewer than 5\% of enterprises have formal quantum-transition plans, and many underestimate "harvest now, decrypt later" risks. Financial, telecom, and government sectors have begun migration, but most industries remain exploratory or stalled by costs, complexity, and skills gaps. Expert consensus places cryptanalytically relevant quantum computers in the 2030s, yet delayed preparation could leave today's data vulnerable for decades. We recommend immediate steps: establishing crypto-agility, creating quantum transition roadmaps, prioritizing PQC deployment in high-value systems, and upskilling cybersecurity teams. A coordinated, proactive approach is essential to secure current and future digital assets in the quantum era.
\end{abstract}


\section{Introduction}
\label{sec:introduction}

Public-key cryptography underpins the security of virtually all digital communications – from HTTPS web traffic and VPN connections to digital signatures and secure email. This security is predicated on mathematical problems (such as integer factorization and discrete logarithms) that are intractable for classical computers \cite{Staat2025,Boyd1993}. However, advances in quantum computing threaten to upend these assumptions. A large-scale quantum computer running Shor’s algorithm could factor RSA keys and break elliptic-curve cryptography, rendering today’s encryption and authentication fundamentally insecure \cite{b5}. The prospect of such a cryptanalytically relevant quantum computer (CRQC) has elevated an urgent question for organizations worldwide: Are enterprises ready for cybersecurity solutions based on quantum computing? This question encompasses both opportunity – the emergence of new quantum-enabled security tools – and threat – the looming “quantum apocalypse” for classical cryptography.

Quantum-secure solutions have emerged in two main forms. The first is post-quantum cryptography (PQC) – new algorithms (still running on classical computers) designed to be resistant to quantum attacks. After a rigorous multi-year process, the U.S. National Institute of Standards and Technology (NIST) has recently standardized a first set of PQC algorithms\cite{Yesina2022}. These include lattice-based key encapsulation (e.g., CRYSTALS-Kyber) and digital signatures (e.g., CRYSTALS-Dilithium and SPHINCS+), which are built on mathematical problems believed to be quantum-resistant. The second category is quantum cryptography, particularly quantum key distribution (QKD), which leverages quantum physics (e.g. the properties of photons) to exchange encryption keys with theoretical information-theoretic security. QKD has seen experimental deployments in specialized networks (for example, financial institutions and government labs) and a growing number of vendors offer QKD products \cite{Aquina2025}. Together, these approaches form the backbone of “quantum cybersecurity solutions” that could protect data against quantum-enabled adversaries.

However, the mere availability of PQC algorithms or QKD devices does not guarantee that enterprises are prepared to implement them \cite{Venkata_Rajesh_Krishna_Adapa_2025,Von_Nethen_2024}. True enterprise readiness spans technological, organizational, and adversarial dimensions. From a technologist’s perspective, one must ask: How mature and practical are quantum-safe solutions today? If PQC algorithms are standardized, are they performant and scalable enough for industry use? Are QKD systems reliable and integrable into existing infrastructure? From a CISO/CIO (enterprise) perspective: Do organizations recognize the quantum threat, and are they making it a priority? This involves budgeting, developing migration plans, managing the complexity of upgrading cryptography \cite{b4}, and overcoming skills gaps. Many enterprises currently face more immediate challenges (such as ransomware, supply chain breaches, and AI-driven threats) that compete for resources and attention \cite{Balzano_2024,Pattnaik_2023}. Finally, from a threat actor perspective: How imminent is the quantum threat in reality, and how are adversaries likely to exploit the transition period? The concept of “store (harvest) now, decrypt later” has already entered the lexicon of cybersecurity agencies \cite{b3}, referring to adversaries hoarding encrypted data today in anticipation of decrypting it once quantum capabilities arrive. Understanding the motives and timelines of nation-states and other threat actors is key to gauging how urgent and extensive the defensive preparations should be.

To tackle these multifaceted questions, this paper employs a Stakeholder Perspective Analysis framework. We analyze recent developments (primarily from 2020–2025) from the three perspectives outlined above – technologists, enterprise security decision-makers, and threat actors – each in a dedicated section. In the Technologist Perspective (Section \ref{sec:2}), we review the state-of-the-art in quantum-safe technologies, including the outcome of the NIST PQC standardization project, performance and deployment considerations of those new algorithms, the status of QKD and its commercial use cases, and the technological readiness levels of these solutions. In the Enterprise CISO/CIO Perspective (Section \ref{sec:3}), we examine studies and surveys on organizational readiness: for example, what percentage of companies have quantum transition strategies, how boards and executives perceive the quantum risk, what challenges (cost, integration, talent) impede action, and which industries are leading or lagging. In the Threat Actor Perspective (Section \ref{sec:4}), we assess the quantum threat landscape: estimates of when quantum computers capable of breaking encryption (“Q-Day”) might realistically materialize, evidence of adversary behavior such as intelligence agencies stockpiling encrypted intercepts, and potential attack vectors including both cryptographic attacks and attacks on quantum systems themselves.

After exploring each perspective, we present a synthesis and discussion in Section \ref{sec:5}. We use a SWOT analysis (Strengths, Weaknesses, Opportunities, Threats) \cite{Leigh_2009} as a tool to integrate the findings: for instance, the strengths might include the progress in PQC standards (an internal positive factor for readiness), whereas weaknesses might highlight internal gaps like the lack of skilled personnel or inadequate budgeting within enterprises. Opportunities could include external positives such as government incentives or new commercial solutions for easing PQC migration, and threats would encompass external negatives like the adversary capabilities and ticking clock of Moore’s Law and quantum research. This structured analysis provides a holistic view of enterprise quantum readiness.

Finally, Section \ref{sec:6} offers a conclusion and recommendations. Rather than a binary verdict (“\textit{yes, enterprises are ready}” or “\textit{no, they are not}”), we present a nuanced conclusion that different sectors and organizations occupy different positions on the readiness spectrum. We also deliver concrete recommendations for action: steps that enterprises should take now to improve their posture, from starting cryptographic inventories and “crypto-agility” projects, to following emerging standards and best practices (such as NIST guidance, or the Quantum Readiness Toolkit from the World Economic Forum) \cite{b17}, to investing in talent development and interdisciplinary collaboration. The goal is to provide a well-supported, multi-perspective assessment that not only answers the research question but also guides stakeholders in preparing for the post-quantum era.

In summary, as we stand at the cusp of a paradigm shift in cybersecurity, this work contributes an up-to-date review and strategic framework for understanding and enhancing enterprise readiness for quantum cybersecurity. The next sections delve into each perspective in detail, leveraging contemporary literature and industry data to illuminate where we stand and what comes next.

\section{The Technologist Perspective: Quantum Cybersecurity Solution Maturity}
\label{sec:2}

From the technologist or solution-provider viewpoint, we assess the maturity, practicality, and scalability of quantum-safe cybersecurity technologies as of mid-2020s. This includes evaluating the readiness of PQC algorithms for deployment and the state of quantum cryptography implementations like QKD. Key questions include: Have viable standards emerged? Are these solutions commercially available and easily integrable? What technical limitations or open challenges remain?

\subsection{Post-Quantum Cryptography Standards and Performance}

The most significant progress in quantum-resistant security has come from post-quantum cryptography \cite{Zhang_2025}. Unlike quantum cryptography, PQC does not require exotic hardware; it consists of classical algorithms (for encryption, key exchange, and digital signatures) designed to be secure against quantum attacks \cite{Deshpande_2024,bbb}. After a worldwide call and competition lasting since 2016, NIST selected its first cohort of PQC algorithms for standardization in July 2022 \cite{b19}. By August 2024, NIST released the finalized standards for three algorithms (designated FIPS 203, 204, 205) covering one key encapsulation mechanism (KEM) and two digital signature schemes. The standardized algorithms are:

\begin{itemize}
    \item CRYSTALS-Kyber (now standardized as ML-KEM) \cite{Nagy_2025,Bos_2018}: A lattice-based key encapsulation mechanism for general encryption/key exchange. Kyber’s appeal includes relatively small key sizes and fast performance, which facilitate practical deployment \cite{b19}. For instance, Kyber public keys on the order of a kilobyte and ciphertexts of a similar size can be transmitted and processed efficiently, making it suitable as a drop-in replacement for RSA or ECC in protocols like TLS.
    \item CRYSTALS-Dilithium (standardized as ML-DSA) and SPHINCS+ (standardized as SLH-DSA) \cite{Liu_2025,Aguilera_2025,Cherkaoui_Dekkaki_2024,Dam2024}: These are quantum-resistant digital signature algorithms. Dilithium is lattice-based, while SPHINCS+ is hash-based (using minimal security assumptions but with larger signature sizes). NIST included both a lattice-based and a hash-based signature in the initial standards, acknowledging the need for diversity in defense (SPHINCS+ serves as a stateless hash-based backup in case a flaw is found in lattice schemes). A fourth algorithm, FALCON (another lattice-based signature), is slated for standardization as an additional option (FIPS 206) by 2025 \cite{Dam2024}.
\end{itemize}

These selections have effectively set global direction – many national agencies and industry groups worldwide are aligning with the NIST choices. Some have also approved certain alternate algorithms for specialized use. For example, German and French cybersecurity agencies (BSI and ANSSI) have signaled acceptance of Classic McEliece (a code-based KEM) and FrodoKEM (an alternate lattice KEM) for conservative long-term security, even though those were not NIST’s primary picks \cite{Aquina2025}. Such endorsements acknowledge that lattice-based schemes, while efficient, are newer and could face unforeseen attacks, so schemes with very different math (like McEliece’s code-based approach) provide hedge against breakthrough attacks.

In terms of security and performance, the standardized PQC algorithms are considered strong against known quantum attack methods (e.g., no efficient attack is known despite extensive cryptanalysis during the NIST process). Nonetheless, they carry certain trade-offs compared to the classical algorithms they replace. Typically, PQC schemes have larger key and signature sizes or higher computational costs \cite{Pandey2023,Hasija2022}. For instance, Dilithium signatures are on the order of a few kilobytes (much larger than an ECDSA signature ~64 bytes), and Kyber’s public key (~800 bytes) is larger than an ECC public key (~32 bytes) but much smaller than some alternatives like McEliece (which uses very large public keys, e.g. $>$ 1 MB) \cite{Raavi_2025,demir2025performance,Jackson_2024}. These differences can impact network bandwidth and storage, especially in constrained devices. However, researchers have demonstrated that for most mainstream applications the overhead is manageable: Kyber operations are fast (comparable to RSA in software and even faster with optimized implementations), and Dilithium can verify thousands of signatures per second on a modern CPU \cite{b19}. Early performance benchmarks indicate that lattice-based algorithms can often be optimized with hardware acceleration (e.g., using vector instructions) and may even outperform classical algorithms at high security levels. Furthermore, many protocols can accommodate larger keys or signatures by negotiating parameters (for example, TLS 1.3 allows larger handshake messages without breaking) \cite{Zheng_2024,Stelzer_2023,Guerrieri_2022}. Thus, from a pure feasibility standpoint, the PQC algorithms are commercially viable. In fact, some applications have already begun trial deployments – for example, Google and Cloudflare have experimented with hybrid TLS configurations (classical + PQC) in Chrome and web servers, and messaging platforms like Signal and Apple iMessage reportedly introduced quantum-resistant encryption modes \cite{b26}.

An important consideration in this context is crypto-agility – the ability to swap out cryptographic algorithms without massive disruption to infrastructure \cite{eee}. The impending transition has underscored that many enterprise systems lack crypto-agility; algorithms are often hard-coded or embedded in hardware, making upgrades laborious. Technologists are now emphasizing designs that decouple cryptographic implementations from business logic, allowing easier updates \cite{cho2024software,eee}. NIST and other bodies have published guidelines for migration to PQC, stressing inventory of all cryptographic use cases and adopting agile frameworks \cite{b17}. The emergence of standard PQC libraries and protocols (for instance, an updated TLS 1.3 or X.509 certificates that support PQC) is accelerating this process. Nevertheless, the maturity level varies: while software libraries (OpenSSL, BoringSSL, etc.) have added PQC support, many hardware devices (smart cards, IoT microcontrollers, older HSMs) currently do not support the new algorithms, implying that technologists face an engineering challenge to achieve parity with the ubiquity of RSA/ECC \cite{b27}.

Another technical challenge is ensuring interoperability and trust in the new algorithms. Standards bodies like the IETF are working on standards for PQC in protocols (e.g., RFCs for PQC in TLS, IPsec, and SSH). Likewise, the WebPKI ecosystem (browsers and Certificate Authorities) needs consensus on how to issue and handle quantum-safe certificates. As of 2025, some Certificate Authorities have begun issuing hybrid certificates (containing both classical and PQC public keys) as an intermediate step \cite{chen2024x,battarbee2024quantum}. These measures are indicative of a transition phase where hybrid cryptography is employed: using both a classical and a post-quantum algorithm in parallel so that security holds unless both are broken. Hybrid modes are intended to provide confidence and backwards compatibility, though they add complexity and must be implemented carefully to avoid new vulnerabilities. There is ongoing research examining potential pitfalls of hybrid approaches (e.g., if one algorithm is weak, does it undermine the combined scheme?) \cite{b28}, but properly designed hybrids can be at least as secure as the stronger of the two algorithms.

It is also noteworthy that post-quantum algorithms are still new cryptographic constructions. This means that unlike RSA and ECC, which have been vetted for decades, some PQC schemes might have undiscovered weaknesses. A cautionary example occurred during the NIST process: the algorithm SIKE (Supersingular Isogeny Key Encapsulation), which reached the final round, was suddenly broken by classical cryptanalysis in 2022 \cite{b29} – a startling result that underscored the need for continued scrutiny. The attack on SIKE did not generalize to other PQC candidates, but it reminds the community that assumptions can be overturned. For this reason, NIST plans a multi-algorithm portfolio and is already working on additional rounds to standardize backup algorithms \cite{Yesina2022}. From a practitioner’s perspective, this implies that technology providers should implement PQC in a modular way, anticipating that algorithms may evolve (somewhat analogous to how hash function transitions occurred after SHA-1’s deprecation). In short, PQC technology is at a Technology Readiness Level (TRL) where standards exist and pilot deployments are underway, but further real-world testing and iterative improvements are expected.

\subsection{Quantum Key Distribution – Capabilities and Limitations}

Parallel to the advances in PQC, Quantum Key Distribution offers a very different approach to secure communications. QKD uses quantum mechanics to exchange encryption keys between parties such that any eavesdropping on the quantum channel can be detected. When combined with one-time-pad encryption or with classical symmetric ciphers, QKD can theoretically provide provably secure communication, independent of an eavesdropper’s computing power \cite{Huang_2024,Asoke_Nath_2024,arxiv.2501.08435}. This promise of “information-theoretic security” is enticing, especially for protecting extremely sensitive data (e.g. government or defense communications with long confidentiality requirements).

Over the past five years, QKD technology has progressed from laboratory experiments to early commercial deployments. For instance, several real-world QKD networks have been demonstrated: financial institutions in Europe and Asia have piloted QKD links to secure bank data backups and inter-bank communications \cite{Aquina2025}; in 2020, the Chinese Micius satellite famously enabled QKD between ground stations over 1,000 km apart, illustrating the potential for long-distance QKD via space-based relays. Companies like ID Quantique, Toshiba, and QuintessenceLabs offer QKD systems that typically operate over optical fiber, with distances up to tens of kilometers (or more with trusted repeaters) \cite{Patel2012,Ali2009}. The global quantum communications market, while still small (valued around \$1.1 billion in 2023), is projected to grow significantly (to \$8.6 billion by 2032) as these technologies mature \cite{Olutimehin2025}.

However, from a technologist’s perspective, QKD in 2025 remains a niche, specialized solution with notable limitations. Major security agencies have scrutinized QKD and pointed out practical issues that currently prevent it from being a general replacement for classical cryptography \cite{Aquina2025,renner2023debate}. The British NCSC, for example, stated in 2020 that it “does not endorse the use of QKD for any government or military applications” and cautions against relying on QKD for business-critical networks, advising instead that “the best mitigation against the threat of quantum computers is quantum-safe cryptography” (i.e., PQC) \cite{Aquina2025}. Similarly, the U.S. NSA has judged quantum-resistant algorithms to be more cost-effective and maintainable, explicitly not supporting QKD for national security systems under current conditions. A joint statement by European agencies (ANSSI, BSI, etc.) concluded that QKD can currently only be used in some niche use cases due to inherent limitations, though research should continue to try to overcome these limits \cite{Aquina2025}.

The technical limitations of QKD that motivate these positions include:

\begin{itemize}
    \item Distance and Infrastructure Requirements: QKD typically requires a direct point-to-point quantum channel (either an optical fiber or free-space optical link) between communicating sites. Optical fiber QKD is usually limited to on the order of 50 - 100 km without a “trusted node” repeater, because photon signals cannot be amplified (amplifiers would disturb the quantum states) and are attenuated by distance. Trusted repeater nodes can extend range, but they must be physically secure (they introduce points that must be trusted implicitly, and could be insider attack targets) \cite{b38}. This fundamentally contrasts with classical encryption which can work over any network path, however indirect, without special relays. Free-space QKD (e.g., satellite) can cover larger distances, but requires line-of-sight and expensive satellite infrastructure. In essence, QKD is not readily scalable to a global network in the way classical cryptography is – it’s best suited for creating “quantum links” in a controlled network environment (e.g., connecting data centers of a single organization) \cite{Pasupuleti2025}.
    
    \item Special-Purpose Hardware: Unlike PQC, which can be deployed via software updates, QKD mandates specialized hardware – single-photon sources, detectors, optical fibers or free-space optics, and often custom key management systems. This means QKD cannot be implemented as a software patch or a cloud service on existing networks \cite{b38}. The need to deploy new physical devices and potentially new fiber links creates high costs. It also means upgrading or patching QKD systems is more involved (you may have to replace hardware to get new features or security improvements). Integrating QKD into existing network equipment is non-trivial; typically, the QKD system runs parallel to classical networks and then keys are fed into encryption devices (e.g., QKD boxes connected to link encryptors). This can be complex to manage and may require integration with classical key management systems \cite{Martin2023,Shokrivahed2024,Lopez2025}.
    
    \item Limited Scope of Security – No Built-in Authentication: QKD solves the problem of distributing a secret key, but it does not by itself authenticate the communicating parties. An adversary can theoretically launch a man-in-the-middle attack on a QKD session if there is no authentication mechanism, by impersonating one endpoint to the other. In practice, QKD links must be bootstrapped with an initial authentication method (usually a pre-shared key or a classical public-key signature to authenticate the first quantum exchange) \cite{b38}. This necessity means QKD is actually used in conjunction with classical cryptography (for authentication), slightly undermining the claim of being purely based on physics – at least some classical trust or pre-shared secret is required. Moreover, if one must manage pre-shared keys for authentication, some ask: why not simply use those keys for encryption as well (e.g., via one-time pads), given the cost of QKD? QKD’s value proposition is strongest when frequent key renewal is needed and high security is required, but it is not a standalone solution – it works as part of a larger cryptographic system \cite{Lai2023}.
    
    \item Security is Implementation-Dependent: While the theoretical security of QKD is proven by quantum physics, real systems have non-idealities. Practical QKD devices have been successfully attacked through various side channels and implementation flaws. For example, researchers demonstrated “blinding” attacks on QKD photon detectors (tricking the system into not noticing eavesdropping), or injecting own signals (Trojan-horse attacks) \cite{b38}. The NSA notes that “\textit{the actual security provided by a QKD system is ... the more limited security that can be achieved by hardware and engineering designs}” \cite{b38} and cites multiple well-publicized attacks on commercial QKD systems. These include timing side channels, pulse-power manipulations, etc., which exploit the fact that real hardware deviates from theoretical models. Such issues mean QKD hardware must be rigorously tested and certified, a process still in nascent stages (there is ongoing work on QKD standards for security certification). Additionally, QKD is highly sensitive to environmental conditions – the same sensitivity that detects an eavesdropper also means QKD links can be easily disrupted by normal disturbances, raising the risk of Denial-of-Service (DoS) \cite{HuguesSalas2018}. A QKD channel might drop whenever a spurious event happens on the line (for instance, extra noise in the fiber), and an adversary could intentionally induce such conditions (by shining light into the fiber, etc.) to block the quantum communication. Thus, while QKD can alert to eavesdropping, it also makes communications fragile: an attacker doesn’t even need to break the encryption; simply jamming the QKD channel can force the system to fall back (or fail) \cite{Maragkopoulos2024}.
    
    \item High Cost and Operational Complexity: Deploying QKD is expensive. It often requires dedicated dark fibers or at least using existing fiber with careful multiplexing and filtering to separate quantum signals from classical ones \cite{Aquina2025}. Running a QKD network may entail hiring specialized quantum engineers for maintenance. Furthermore, most use cases require not just a point-to-point link but a network. Building a QKD network at scale (with many nodes) leads to a combinatorial explosion of needed links or introduction of intermediate trusted nodes \cite{Chen2021}. Some network architectures like a “star topology” with a central trusted node can distribute keys to many endpoints, but then that node becomes a critical trust center. Overall, current QKD is not easily scalable to large user bases – its Technology Readiness Level might be considered adequate for pilot systems and niche deployments but not for broad enterprise adoption across arbitrary distances.
    
\end{itemize}

Given these challenges, it is widely recognized by technologists that post-quantum cryptography is the primary path for most cybersecurity applications, whereas QKD is a complementary solution for specialized scenarios. PQC can be deployed in software on existing infrastructure and provides broad protection (e.g., one can upgrade web browsers, servers, VPNs, and IoT devices with PQC algorithms). By contrast, QKD might be justified for securing certain fiber-optic links that carry extremely sensitive data (like between data centers or between a bank’s headquarters and a backup site), where the added security is worth the cost and where the distance is within feasible limits \cite{Aquina2025}. In such cases, QKD would typically be used alongside classical encryption (often referred to as a “hybrid” solution: keys from QKD feed a symmetric cipher like AES, while classical PQC or pre-shared keys ensure authentication).

Notably, research continues to push QKD forward: efforts are underway to integrate QKD with telecom networks (e.g., using wavelength-division multiplexing to send quantum and classical signals in the same fiber) \cite{Aquina2025}, to improve distances via quantum repeaters (not yet practical, but quantum memory and entanglement swapping might extend range without trusted nodes in the future), and to develop standardized QKD protocols and certification criteria. Also, Quantum Random Number Generators (QRNGs) – a related technology – are already being used to bolster cryptographic systems by providing truly unpredictable keys, an example of quantum technology immediately useful within classical cryptography frameworks.

In summary, the technologist perspective finds that quantum-safe solutions are emerging from concept to reality. Post-quantum cryptography is at the forefront, with multiple algorithms standardized and early implementations proving feasible for real-world use \cite{Yesina2022}. While some performance and integration challenges exist (e.g., handling larger keys, updating protocols), these are being actively addressed, and major technology providers are incorporating PQC support. Quantum key distribution, on the other hand, remains a more specialized tool – it has reached the stage of operational prototypes and limited real-world use, but with substantive constraints (distance, cost, and trust issues) that confine its applicability. The consensus among experts is that PQC will be the workhorse of future quantum-safe security, whereas QKD might protect specific high-value links where its unique advantages outweigh its drawbacks \cite{Aquina2025}. This perspective sets the stage: the tools for quantum-safe cybersecurity exist or are rapidly maturing. The next question is whether enterprises are in a position to adopt them, which we examine from the CISO/CIO viewpoint.

\section{The Enterprise Perspective: Evaluating Business and Operational Readiness}

\label{sec:3}

From the enterprise decision-maker’s perspective (chief information security officers, CIOs, risk managers), we evaluate how prepared organizations are to embrace quantum-safe cybersecurity. This includes awareness of the quantum threat, prioritization relative to other business risks, resource allocation (budget and personnel), practical challenges in integration (legacy systems, operational disruption), and any early adoption efforts. We rely on industry surveys, reports, and case studies to gauge the current state of enterprise readiness in 2024–2025.

\subsection{Awareness vs. Action: The Quantum Risk on the Executive Radar}

In recent years, awareness of quantum computing’s security implications has grown among business and security leaders, yet this awareness has not consistently translated into tangible preparation. Multiple surveys reveal a stark readiness gap. For example, a global poll by ISACA in 2025 found that 62\% of technology professionals are worried that quantum computing will break current encryption – a clear majority acknowledging the risk – but only 5\% of organizations have made it a high-priority issue in the near term \cite{b1}. In the same survey, just 5\% reported having a defined quantum computing strategy or roadmap in place. In other words, fewer than one in twenty enterprises have a plan to address quantum threats, despite widespread concern that core security protocols could become obsolete.

This disconnect arises partly from human nature and business incentives: quantum threats have long been portrayed as “future problems” with uncertainty about timing. Many executives adopt a “wait and see” stance, especially when more immediate crises abound. In 2024, topics like ransomware, supply chain attacks, and AI-driven threats (e.g., deepfakes, generative AI misuse) dominated CISO agendas, often overshadowing longer-horizon issues. Indeed, in $ISC^2$ ’s 2024 global security workforce study, quantum computing was ranked the second most concerning emerging technology risk (after AI) \cite{b53}, yet pressing day-to-day concerns make it challenging to justify budget and staffing for an issue that might not fully manifest for 5–10 years. One result is procrastination; as an ISACA analyst quipped, “only 5 percent of respondents consider addressing quantum threats a high business priority for the near future” \cite{b54}, indicating that for 95\% of firms, it’s medium or low priority at best.

However, some sectors are notably more proactive. A detailed industry study by Capgemini Research Institute (2025) revealed that in sectors handling highly sensitive data or facing strict regulations, many organizations are starting the transition to PQC much earlier than the broader market. For instance, the survey of “early adopters” found that nearly two in five organizations planned to initiate their PQC transition in the next two years (by 2025–2026). Breaking it down by industry: banking and telecommunications were leading, with ~45–47\% of respondents in those sectors saying they have budgeted and planned for PQC in the near term \cite{b7}. Defense and high-tech industries followed close behind at ~43\%. These are sectors where the risk of being unprepared is unacceptably high – banks, for example, not only have long-lived financial data and obligations to secure customer information, but also are likely targets of nation-state hackers who might employ quantum capabilities as soon as they can. Likewise, telecom operators form the backbone of secure communications and are critical infrastructure, so their incentive to safeguard against future threats is strong \cite{b7}.

In contrast, sectors like consumer products, retail, and some areas of manufacturing showed much lower immediate engagement with PQC (often focusing on other digital transformation initiatives unless they have specific compliance drivers). Many such c
ompanies report that they “currently see no business case for PQC” or have “no timeline yet; still in exploration”. In the Capgemini data, a significant percentage of organizations admitted they have not started any quantum-security planning because it’s either not a priority or they are waiting for clearer mandates \cite{b7}. In fact, 41\% of organizations said they do not plan to address quantum computing at this time, and 37\% have not even discussed it internally according to the ISACA 2025 poll \cite{b1}. These numbers underscore that outside of a motivated minority, enterprise action lags far behind the technology curve.

One driver that can shift an issue from awareness to action is regulatory or government pressure. In the case of quantum readiness, we are beginning to see such pressures emerge. The U.S. government, through National Security Memorandum 10 (NSM-10) in 2022, has set requirements for federal agencies to inventory their cryptographic systems and plan for a transition to PQC, with a goal of mitigating as much quantum risk as possible by 2035 \cite{b5}. As part of this, agencies and their suppliers are being nudged to start implementing quantum-safe solutions in the mid-2020s. Similarly, the U.S. Office of Management and Budget issued guidance (OMB Memo 22-09 and 23-02) with deadlines for federal agencies to begin encryption inventory and migration processes. These government mandates indirectly influence enterprises, especially those in the supply chain for government or critical infrastructure. In Europe, while no overarching regulation compels PQC adoption yet, there is growing attention from cybersecurity authorities and the EU’s cybersecurity agency (ENISA) about quantum threats, and we may see future standards or directives. The United Kingdom’s NCSC has recommended that organizations have an initial PQC migration plan by 2028 and begin their highest-priority migrations by 2030–2031 \cite{b7}. Such guidance essentially puts a backstop on procrastination – it signals that by the late 2020s, not having a plan will be deemed irresponsible.

Another factor is client and market expectations: as large enterprises (e.g., global banks, cloud providers) start boasting of “quantum-resistant security” in their offerings, it may spur others to follow to avoid being seen as the weak link. Already, some technology service providers use quantum-resilience as a selling point; for example, certain VPN and secure communications products advertise that they are “quantum-safe” by integrating PQC, and cloud companies like AWS, IBM, and Microsoft have released tools for experimenting with PQC in their platforms.

\subsection{Key Challenges in Implementation and Integration}

Even once an enterprise decides to act on quantum security, they encounter several practical challenges that can slow progress. Key among these are: (a) the complexity of identifying and upgrading legacy systems, (b) resource constraints (both budget and skilled personnel), (c) balancing quantum-security investments with other immediate priorities, and (d) uncertainty in standards and interoperability.

\textbf{Legacy Infrastructure and Technical Debt:} Large organizations might have hundreds of applications and systems that rely on cryptography – from public-facing websites and VPNs to internal software, databases, and IoT devices \cite{Kuzminykh2020}. Many of these systems use hard-coded cryptographic algorithms. A major challenge is simply discovering everywhere that vulnerable cryptography (RSA, ECC, DH, etc.) is used – a process often called a cryptographic inventory. This is non-trivial; some legacy applications might not even have up-to-date documentation. Once identified, each instance must be assessed for upgradability: Is the algorithm baked into hardware (as in many IoT or OT devices)? Can the software be patched? Will upgrading to a PQC algorithm require a more powerful processor or more memory than the device has? These issues contribute to what might be called crypto technical debt in organizations. As an example, consider the use of encryption in embedded industrial controllers or medical devices – if those devices are still in use and cannot be easily replaced or patched, they represent a risk of becoming “stranded” with insecure crypto. A 2024 analysis by Forescout’s Vedere Labs found that 75\% of OpenSSH instances in use (on internet-facing systems) were running versions from 2015–2022 that do not support quantum-safe algorithms, and less than 20\% of TLS servers had been upgraded to TLS 1.3 (the only version of TLS designed to easily accommodate PQC) \cite{b9}. This highlights how much existing tech is outdated in terms of crypto-agility. Upgrading such a wide swath of infrastructure will be a slow, meticulous process, often 
aligned with regular tech refresh cycles. Enterprises thus face a planning challenge: how to incorporate PQC updates into their roadmap, potentially prioritizing systems that protect the most sensitive data or have the longest expected lifetime.

\textbf{Resource and Talent Constraints:} Budget is a perennial challenge. A quantum-safe overhaul may not show immediate ROI (it is fundamentally a risk mitigation spend). Many organizations operate on tight IT and security budgets where an increase of a few percentage points is hard-won. According to industry research, security budgets in 2024 were growing around 6–8\% on average \cite{b62}, which often just keeps up with inflation and the rising costs of conventional threats. Carving out a portion of that for quantum preparedness means convincing senior management of the long-term benefit. Some forward-looking organizations have done that; others have not. Talent is an even more acute issue: cryptography is a specialized field, and post-quantum cryptography is newer still. Most security teams do not have a cryptographer on staff. In a 2025 survey, 68\% of organizations reported struggling to find or develop the skills needed for quantum-safe implementations \cite{b7}. Over half (61\%) also cited the lack of clear industry guidelines or standards as a concern, and 49\% worried about regulatory uncertainties in this area \cite{b7}. The shortage of qualified experts means that enterprises often must rely on external consultants or vendors for guidance, which can be costly and creates dependence. Some financial companies have started dedicated internal “quantum risk” working groups, essentially upskilling some of their cryptography and risk personnel, but these remain the exception. ISACA’s poll found that 52\% of respondents believe quantum will change the skills needed in their business (e.g., needing more mathematicians, quantum-savvy engineers) \cite{b1}, yet few have concrete hiring or training plans in place for that shift. The recommendations from analysts are that organizations begin capacity building now – e.g., sponsoring current engineers to take courses in PQC, partnering with universities or consortia on pilot projects \cite{b7}. In practice, however, with the overall cybersecurity talent gap already large (the global workforce gap is estimated in the millions \cite{b53}), adding quantum requirements exacerbates the challenge.

\textbf{Competing Priorities and Risk Perception: }Enterprises constantly juggle risks – cyber and otherwise – trying to allocate resources where they are most needed. Quantum risk often suffers from being perceived as less immediate than things like zero-day vulnerabilities or compliance deadlines for regulations such as GDPR or new privacy laws. Many CISOs take a probabilistic risk management view: if the consensus is “CRQC by ~2035,” they might reason that there is time and that scarce budget this year should go to augmenting an understaffed SOC or buying an improved XDR solution to stop current attacks \cite{Battaglioni2022}. This view is risky if it leads to deferral beyond the point of no return. Yet, it’s important to note that some organizations are explicitly using quantum readiness as a lever to secure bigger budgets now – turning the issue on its head. Forward-leaning CISOs have argued to their boards that quantum preparedness is part of future-proofing and compliance (since regulators are starting to mention it). By framing it as an upcoming compliance requirement (“\textit{we will likely be mandated to do this soon, so starting early avoids future penalties or scrambles}”), they manage to get buy-in \cite{b70, b71}. This tactic is mentioned in some advisory blogs: tying quantum risk mitigation to overall risk tolerance and using it to justify budget increases for encryption modernization \cite{b72, b73}. Still, for many companies, until there is an actual mandate or a well-publicized quantum breakthrough that shocks the industry, getting sustained executive attention is difficult. The ISACA 2025 report's headline – “\textit{Despite Rising Concerns, 95\% Computing Roadmap}” – encapsulates this current state \cite{b1}.

\textbf{Uncertainty in Standards and Technology Evolution:} While NIST has provided certainty by standardizing some algorithms, there is still some unsettled ground that enterprises cite as reasons to wait. For example, some organizations worry: What if the algorithms change again? – They recall the tumult when certain encryption standards (like SHA-1) were deprecated and replacements had to be rolled out. The reality is that NIST’s PQC standards are expected to be stable (they underwent years of vetting), but the ongoing process for additional algorithms means more standards will come by 2024–2025 (for instance, another KEM or signature) \cite{Yesina2022,lamacchia2025post}. A prudent strategy is not to let this halt progress, but to adopt a flexible approach. In practice, companies can start by implementing one of the approved algorithms (e.g., Kyber and Dilithium) in a crypto-agile manner, so that if a new standard or a required change emerges (say a vulnerability in one algorithm), they can pivot. Nonetheless, the lack of widely adopted protocols and tools for some use cases is a challenge. Until very recently, if a company wanted to deploy PQC in say, their VPN, they had to use specialized or experimental software. This is improving: major VPN vendors and TLS libraries now have beta support for PQC. The ecosystem is catching up – e.g., OpenSSL 3.0+ includes support for certain PQC algorithms, and new versions of programming languages and libraries are starting to bundle in PQC as optional \cite{ahmed2025survey,demir2025performance}. Still, early adopters often have to do extra integration testing, and enterprises that lack strong R\&D divisions may prefer to wait for more off-the-shelf solutions. A report noted that many organizations plan to rely on their vendors to “bake in” quantum safety once it’s ready; in other words, they expect cloud service providers, software vendors (databases, operating systems), and network appliance vendors to handle the heavy lifting, delivering updates that the enterprise can then just apply \cite{bishwas2024strategic}. This expectation is reasonable to a degree – indeed, cloud and software vendors are working on it – but the timeline is uncertain. If all users wait for vendors and all vendors wait for clear demand from users, it creates a stalemate. Fortunately, the largest tech companies are not waiting – they are deeply involved in PQC and even QKD research and incorporating them into their offerings (IBM for instance has a quantum-safe cryptography service for mainframes, and AWS offers a toolkit to scan and test cryptographic usage) \cite{Nther2024}.

\textbf{Compliance and Legal Considerations:} Another aspect of readiness is whether adopting (or not adopting) quantum-safe measures ties into compliance requirements or liabilities. As of 2025, no major regulation explicitly penalizes a company for not using PQC. However, there is a possibility in the future of regulations in finance, healthcare, or critical infrastructure that require quantum risk assessments. Already, sectoral regulators are asking questions: e.g., banking regulators have started including queries about “crypto-agility plans” in their exams for large banks, and the U.S. FDA has signaled that medical devices being approved in late 2020s should consider crypto-agility (since those devices might remain in use into the 2030s) \cite{Ogundola2025}. An organization that fails to anticipate these could find itself needing a crash program later. Some legal experts also warn of a scenario: if a breach happens in, say, 2032, and encrypted data is stolen and then later decrypted by an attacker using a quantum computer, could an organization be considered negligent (in hindsight) for not protecting long-lived sensitive data with quantum-resistant measures? It’s a grey area, but as the timeline clarifies, what is “reasonable” security practice will evolve.

In summary, the enterprise perspective at present can be characterized by slow mobilization with pockets of leadership. Most enterprises are in early stages – discussions or explorations – rather than execution of quantum-safe migrations. The ones taking action are often either in highly regulated or sensitive sectors or have visionary leadership that treats quantum
risk as an opportunity to strengthen overall crypto governance (performing a long-needed “cryptographic hygiene” overhaul). Key weaknesses hindering readiness include the lack of skilled personnel, limited budgets specifically earmarked for this purpose, and inertia/other priorities (51\% of organizations cited “organizational inertia” as a barrier in one survey \cite{b7}). The strengths that some organizations possess include strong governance (companies that already have a mature cryptography management practice find it easier to slot in PQC), support from top executives who understand the strategic risk, and involvement in industry collaboration (those participating in standards groups or pilot programs gain knowledge faster).

A crucial insight from this perspective is that the timeline to act, from an enterprise change-management view, is actually quite short. Upgrading cryptography across a large enterprise can easily take 5–10 years (when considering budgeting, procurement, development, testing, and rollout across thousands of systems). Thus, even if a working quantum computer is a decade away, starting now is not premature. Many experts and agencies explicitly urge starting immediately \cite{b5}. For example, Dustin Moody (NIST PQC project lead) said in 2024: “We encourage system administrators to start integrating (the new standards) into their systems immediately, because full integration will take time” \cite{Yesina2022}. The U.K. NCSC’s roadmap of plan by 2028 and earliest migrations by 2030 aligns with that notion of ~5-year lead time for major changes \cite{b7}. Unfortunately, as data shows, the majority of enterprises have not yet internalized this urgency, which could leave them scrambling later.

Before turning to what can be done about this (in our recommendations), we will incorporate the third perspective – that of the threat actors – which often provides the “push” needed for enterprises to act. Seeing how adversaries are viewing and preparing for the advent of quantum computing can serve as a wake-up call for business leaders who might otherwise be complacent.

\section{The Threat Actor Perspective: Characterizing the Quantum Threat Landscape}
\label{sec:4}

This section examines the quantum computing threat from the point of view of potential adversaries – from cybercriminal groups to nation-state intelligence agencies. We address questions such as: How imminent is the “quantum threat” really? What do experts believe about the timeline for a CRQC? Are threat actors already taking steps (like collecting encrypted data now) in anticipation of future decryption? What kinds of attacks or targets are of most interest? Additionally, we note that threat actors could target not only classical cryptography (via quantum algorithms like Shor’s) but also any new technologies (for example, attempting to attack QKD systems or PQC implementations via side-channels or classical means). This perspective underscores the urgency and informs which areas of defense need priority.

\subsection{Timeline to “Q-Day” and Adversary Capabilities}

One of the most pivotal factors is when a quantum computer capable of breaking current cryptography might become operational – often dubbed “Q-Day”. There is a range of views on this, but a common thread is uncertainty. A National Academies study and other expert assessments have often been summarized as saying it is unlikely to happen before the 2030s \cite{b5}. For instance, in late 2023, RAND Corporation noted an “expert consensus” that CRQCs will not be developed until at least the 2030s. The reasoning is that the technical challenges to build a large-scale, error-corrected quantum computer are formidable – requiring perhaps millions of physical qubits to get a few thousand stable logical qubits, enormous error-correction overhead, and breakthroughs in hardware engineering. On the other hand, more optimistic projections have emerged from some quantum computing companies and academics, suggesting a 50/50 chance of such a machine by, say, 2030. A 2025 Cyber Defense Magazine article 
highlighted that some experts forecast a quantum computer able to break RSA-2048 could exist in as little as 5–10 years (by ~2029–2030), while others estimate it may take until 2034 or even 2044 \cite{b4}. In other words, credible voices can be found for both an aggressive and a conservative timeline.

Given this uncertainty, many security agencies adopt a prudent stance: prepare for the earliest reasonable breakthrough, because the cost of being caught unprepared is catastrophic. The U.S. NSA, for example, has stated that the “adversarial use of a quantum computer could be devastating to national security systems” \cite{b5}, and thus they assume a defensive posture that such a capability could appear during the 2030s or perhaps sooner. The NSA’s timeline for transitioning its own systems (via the Commercial National Security Algorithm Suite 2.0) reflects an aim to be mostly quantum-safe by around 2035. Similarly, the U.S. White House NSM-10 set 2035 as a goal to mitigate most quantum risk \cite{b5}. These timelines implicitly indicate a belief that a CRQC is unlikely much before then, but not so late as 2050 – essentially hedging for mid-2030s. The European Telecommunications Standards Institute (ETSI) Quantum-Safe group and other international bodies likewise often mention “10–15 years” as an estimation, though always with a caveat that progress is not entirely predictable.

Crucially, adversaries will not announce when they have a cryptographic-breaking quantum computer – if a nation-state achieves it, it would almost certainly be a closely guarded secret initially. However, it is also widely argued that building such a machine in secret without tipping off the scientific community would be extraordinarily difficult \cite{b5}. Quantum computing research is global and involves hundreds of researchers; breakthroughs (and even incremental progress) tend to be published. Edward Parker (RAND) noted that “it is extremely unlikely that any organization will develop a CRQC in secret” due to the need for large engineering efforts and the fact that any significant advances (like notable qubit records, etc.) are hard to hide \cite{b5}. Thus, completely covert development is improbable – we are likely to see warning signs (e.g., a sudden jump in published qubit counts, or certain milestones like demonstration of error-corrected logical qubits, etc.). Even intelligence agencies rely on the broader R\&D ecosystem, which leaks information.

Nonetheless, from a threat perspective, one cannot rely on perfect transparency. The worst-case scenario – often dubbed the “black swan” or nightmare scenario – is if a hostile actor somehow got a CRQC early and used it covertly for years before the world realizes. In such a scenario, they could quietly decrypt vast amounts of intercepted traffic and harvested data without tipping off victims. As discussed by RAND and others, this scenario is deemed highly unlikely for multiple reasons \cite{b5}. But even if a CRQC’s existence is not secret, there’s a worry that once it is plausible that someone has it, panic could ensue among defenders trying to rapidly upgrade systems (“crypto scramble”).

Thus, threat actors (particularly nation-states) are likely strategizing for both eventualities: the pre-Q-Day and post-Q-Day phases. In the pre-Q-Day phase (present to ~Q-Day), the primary threat actor behavior is the “Store/Harvest Now, Decrypt Later” (HNDL) approach. As soon as it became broadly believed that quantum computers will eventually break RSA/ECC, intelligence agencies and even some cybercriminal groups realized that any encrypted data stolen today might be readable in the future. So why not steal as much as possible now and archive it? Indeed, western security agencies have explicitly warned that “adversaries may be collecting encrypted communications now, anticipating that quantum computers will eventually break current encryption standards” \cite{b9}. The U.S. Cybersecurity and Infrastructure Security Agency (CISA) and the UK’s GCHQ have echoed this warning. A dramatic example came to light in mid-2024: The Volt Typhoon cyber-espionage campaign (attributed to Chinese state actors) was found to have infiltrated telecom and infrastructure networks, exfiltrating large amounts of data including encrypted communications. While there’s no direct evidence they have a decryption capability yet, the implication is that they were collecting encrypted data for long-term exploitation. Types of data likely targeted include diplomatic cables, military communications, intellectual property (trade secrets, proprietary formulas/designs), and personal data on high-value individuals – basically, anything that could retain its importance for years. A survey of security professionals by ISACA found that 56\% were specifically concerned about the HNDL tactic (the fact that encrypted data stolen today could be broken later) \cite{b1}. This indicates not just an abstract fear, but that organizations are aware their current intercepted traffic or breaches could come back to haunt them if not quantum-protected.

What about non-nation-state actors? Cybercriminals motivated by profit have less incentive to store data for a decade (stolen credit card numbers won’t be useful by then, for example). Their interest in quantum would more likely be once a quantum breaking capability exists, they might use it opportunistically. But some criminal organizations with ties to states or extremely patient investors might archive certain things (e.g., encrypted wallet credentials or financial transaction records that could yield money if cracked later). The primary HNDL actors are likely state-sponsored APTs focusing on strategic intelligence.

In the post-Q-Day phase, if and when a CRQC comes into existence, we can expect a rapid shift in the threat landscape. Anyone with access to such a machine (initially a few great powers, perhaps) could essentially undermine the security of most public-key based protections on the internet: TLS handshakes, VPN tunnels, SSH, code-signing certificates, blockchain signatures, etc. The threat actor’s “dream” is to retroactively unlock years of confidential data or to impersonate others by forging signatures (e.g., fake software updates). A major concern is critical infrastructure: for instance, secure SCADA communications or over-the-air updates to devices might be targeted. If one country first achieves CRQC, it might use it surreptitiously to advantage – e.g., reading another country’s classified communications until the victim upgrades to PQC. Once the capability is believed to exist, all organizations globally would face pressure to immediately transition, likely causing chaos for those unprepared \cite{b5}. Indeed, RAND’s analysis argued that as soon as it’s known that a CRQC is plausible, most organizations will “immediately move to upgrade” to PQC, albeit in a potentially “expensive, chaotic, and disruptive” manner. This is essentially the doomsday scenario that defenders want to avoid by acting early.

It’s also worth mentioning cryptanalysis and alternative attacks: threat actors might not necessarily wait purely for a quantum computer. They could also focus on breaking the new PQC algorithms or exploiting weaknesses during the migration. For example, a concern is that in the rush to implement hybrid solutions (say combining RSA and Kyber in a TLS handshake), misconfigurations or subtle protocol flaws might occur, giving attackers a loophole. If organizations run “dual stack” systems (supporting both old and new crypto), attackers will target the weaker link (perhaps forcing a downgrade to the old algorithm if possible, akin to how TLS downgrade attacks work). Nation-state attackers could also invest in classical cryptanalytic advances in the interim; one lesson from the PQC selection was that a couple of the candidate algorithms were broken by classical means (e.g., the case of SIKE in 2022) \cite{b29}. Adversaries will surely use both classical and quantum means: e.g., if one of the NIST PQC algorithms later shows a vulnerability, attackers will jump on it.

Threats to Quantum Solutions Themselves: It’s not only classical systems at risk; if organizations deploy QKD or other quantum tech, those can be targeted too, albeit by different methods. As noted earlier, QKD systems have seen “quantum hacking” demonstrations where eavesdroppers exploit hardware loopholes. A capable adversary could replicate these if, say, a government or bank relies on QKD. Moreover, disrupting quantum comms via jamming is a likely tactic since it’s easier than breaking the physics. We should also consider supply chain threats: the security of PQC depends on correct implementation, so malicious actors might attempt to introduce weaknesses in software libraries or hardware accelerators that implement PQC (for instance, a backdoored random number generator could undermine even a strong algorithm). This is analogous to past events (like the Dual\_EC RNG backdoor incident) – adversaries will still play those games in the PQC era.

Another dimension is the “Y2Q” problem – a play on Y2K, referring to the moment when quantum breaks hit. Just as Y2K required remediation of date handling, Y2Q requires remediation of cryptography \cite{Zhang2023,Zhang2021}. Threat actors in the lead-up might attempt to accelerate their operations to take advantage of any window of insecurity. Conversely, after PQC is deployed widely, some older threat tools (like typical phishing that relies on tricking users to install malware) won’t change; quantum doesn’t fix everything, it’s a specific threat to cryptography.

In summary, the threat actor perspective reinforces that the threat is credible and growing. The timeline might be uncertain, but adversaries are not sitting idle. Nation-state attackers are almost certainly investing heavily in quantum R\&D themselves – China, for instance, has publicized major investments in quantum computing and quantum communications, with strategic documents indicating their intent to lead in these fields. We can infer these states aim to both use quantum tech for offense (decryption) and defense (their own quantum-safe comms). The existence of programs like China’s 2,000-km QKD network and quantum satellites shows they take the future threat seriously (even if one could argue QKD is more a strategic hedge). Western agencies like NSA and GCHQ, by pushing PQC adoption urgently, reveal their belief that adversaries (read: possibly Russia, China, others) will exploit any lag in upgrading.

A key threat identified here – the harvesting of encrypted data – means that for data with a long shelf-life, the effective risk is already present today. If an enterprise has, for example, sensitive trade secrets or customer PII that must remain confidential for 10+ years, then that data, if intercepted now, is under threat even if Q-Day is 10 years away. This flips the problem from a future one to a present one, at least for certain classes of information \cite{halak2024security,Dima2011}.

Thus, from a threat perspective, the urgency is real: whether Q-Day is 8 years or 18 years away, actions by adversaries today (harvesting intelligence) and the strategic ramifications of a sudden quantum leap justify treating this as a current strategic risk. The “window to act” before the threat materializes is closing year by year \cite{b9}. In the next section, we synthesize the findings of all three perspectives – technology, enterprise, and threat – to evaluate overall readiness and to identify how enterprises can strengthen their posture against this looming quantum threat.

\section{Synthesis and SWOT Analysis of Enterprise Quantum-Readiness}
\label{sec:5}

\begin{figure}
    \centering
    \includegraphics[width=0.8\textwidth]{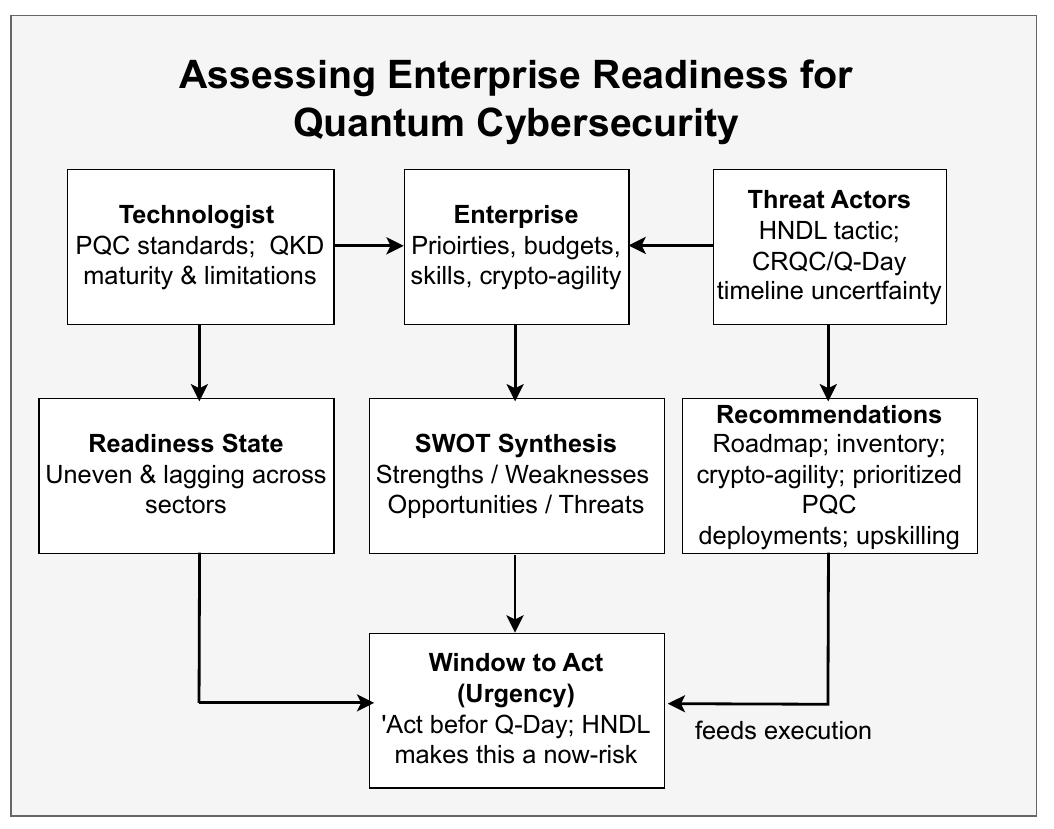}
\caption{SWOT analysis of enterprise readiness for quantum-safe cybersecurity, highlighting that while technology maturity presents strengths and opportunities, internal weaknesses and external threats require urgent action.}
    \label{fig:fig1}
\end{figure}

Bringing together the three perspectives above yields a comprehensive but complex picture. It is evident that technologically, viable solutions are either available now or will be soon (PQC standards are finalized, and even QKD is serviceable in niches), and adversaries are motivated and preparing, yet most enterprises lag in actual preparedness. To better structure this analysis and derive strategic insights, we employ a SWOT framework (Fig. \ref{fig:fig1}), summarizing the Strengths, Weaknesses, Opportunities, and Threats related to enterprise readiness for quantum cybersecurity. This SWOT analysis spans the internal factors (strengths and weaknesses within enterprises or inherent to the technologies) and external factors (opportunities and threats arising from the broader environment, including adversary behavior and market/regulatory trends).

\begin{itemize}
    \item \textbf{Strengths (Internal Positive Factors):} These include encouraging elements within the current situation that enterprises can leverage. One key strength is the head start in technology development – the world is not entering the quantum era empty-handed. The development of post-quantum cryptography standards is a major strength; enterprises now have a menu of vetted algorithms (lattice and hash based) that they can start deploying, backed by extensive academic review \cite{Yesina2022}. This means the defense is not theoretical – practical tools exist, and moreover, many current security protocols (VPNs, TLS, etc.) have already been experimentally shown to work with PQC with acceptable performance. Another strength is that quantum risk has begun to register on the radar of corporate boards and executives (thanks in part to awareness efforts by government agencies). The fact that 62\% of professionals are worried about encryption-breaking \cite{b1} suggests that at least within security teams, the issue is understood, which is the first step to action. Some large enterprises have turned this awareness into proactive initiatives: for example, several big banks and telecoms have internal working groups and have conducted successful pilot projects for PQC integration \cite{b7}. Those early adopters create a knowledge base and possibly set industry examples that others can follow. Additionally, existing security processes can be adapted for quantum preparations: many organizations already have risk management frameworks (like ISO 27001, NIST CSF) – integrating quantum risks into those (e.g., updating risk register, conducting quantum risk assessment as part of enterprise risk management) is straightforward and starting to happen. On a technical level, symmetric cryptography and one-way hash functions remain strong (doubling key lengths can counter Grover’s algorithm) \cite{b29}, so enterprises don’t have to overhaul everything – their VPN’s AES encryption, for example, is still safe as long as keys are long; this is a strength because it narrows the problem mainly to public-key infrastructure. Lastly, global collaboration is a strength: unlike some security problems that each enterprise must face alone, the quantum threat has galvanized a collective response – open global standards, cross-industry consortia (like the Internet Engineering Task Force PQC Working Group, the Global Quantum Safe Forum, etc.), and public-private partnerships (e.g., NIST working with industry on migration guidelines). This means enterprises, even if they lack internal expertise, can draw on a rich pool of external knowledge and tools.
    
    \item \textbf{Weaknesses (Internal Negative Factors):} These represent the current shortcomings and obstacles within enterprises that hinder readiness. Foremost is lack of prioritized action and strategies: as evidenced, 95\% of organizations lack a formal roadmap \cite{b1}. Many companies have not allocated budget or assigned responsibility for this issue – it tends to fall in the cracks between IT, security architecture, and risk management. This organizational inertia is a serious weakness. Another weakness is shortage of talent and expertise: few security teams have members deeply familiar with PQC algorithms or implementation pitfalls. The learning curve can be steep, and not enough is being done yet to train staff (though there is recognition of this gap \cite{b7}). Relatedly, crypto-agility is lacking in many existing systems; years of using a stable set of algorithms (RSA/ECC) led to complacency in software design. Enterprises now find that upgrading algorithms might require software refactoring or even hardware replacement. Legacy technology debt – old systems that cannot be easily updated – is a critical weakness, meaning some businesses will have to maintain insecure algorithms in use longer than they’d like because, say, a critical piece of manufacturing equipment only supports a certain protocol. Another weakness lies in unclear risk quantification: many organizations have not translated the quantum threat into concrete risk metrics or dollar impacts. When risk is not quantified, it’s harder to justify investment – hence the business case is perceived as weak in many quarters. Some surveys show that a significant fraction of enterprises “see no business case” for PQC at present \cite{b7}. This is a weakness of communication and perspective: those organizations likely undervalue the long-term cost of inaction. Additionally, competing priorities and cyber fatigue within security teams mean that quantum projects often get deferred; there is only so much a team can handle alongside urgent incidents and regulatory compliance tasks. Lastly, on the tech side, while PQC algorithms exist, tooling and support are still immature in many areas – e.g., lack of FIPS-validated cryptographic modules for PQC (though they will come), lack of turnkey solutions. This can make early deployments error-prone (weakness in reliable implementation).
    
    \item \textbf{Opportunities (External Positive Factors):} These are emerging or future factors in the environment that enterprises can capitalize on to improve readiness. One major opportunity is the acceleration of solutions and services in the market: a number of cybersecurity vendors are now offering “quantum-safe” options – for example, PKI vendors providing certificates that use PQC, VPN/appliance vendors with quantum-safe modes, and cloud providers integrating PQC into their crypto services. As these offerings mature, enterprises have the opportunity to adopt quantum safety without bearing the full R\&D cost themselves. Another opportunity is shaped by regulatory foresight: organizations that act early will not only reduce risk but possibly gain favor with regulators and customers for being forward-thinking. For instance, banks that implement PQC ahead of regulatory mandates could advertise enhanced security for client data, potentially gaining a competitive edge (especially in sectors like finance or healthcare where customers are sensitive about long-term confidentiality, or government contracting where quantum-safe capabilities might become a procurement criterion). There’s also an opportunity for funding and support: government initiatives (like grants for quantum readiness, or tax incentives for upgrading critical infrastructure security) may become available. Already, national initiatives such as the U.S. National Quantum Initiative and EU’s Quantum Flagship, while primarily research-focused, create ecosystems that companies can join, sometimes benefiting from shared pilot projects or testbeds \cite{Aquina2025}. Inter-industry collaboration is an opportunity – enterprises can join consortiums or working groups to share knowledge on migration strategies (for example, the Bank of England convened a Quantum-Safe Financial Services group). By working together, enterprises can lower costs and avoid duplication of effort. On the technology front, improving automation and scanning tools provide opportunity: emerging tools can scan codebases to find hard-coded algorithms or keys, accelerating the inventory process. The rise of devops and continuous integration means once an organization decides to make systems crypto-agile, they can potentially deploy changes faster than in the past. Finally, opportunity in incident avoidance: being proactive on quantum security not only averts the quantum-specific disaster, it likely strengthens overall crypto discipline (for example, an inventory might uncover forgotten weak crypto like 1024-bit RSA still in use, which can be fixed now). So quantum preparedness efforts often yield side benefits of better classical security – an opportunity to bolster defenses holistically.
    
    \item \textbf{Threats (External Negative Factors):} These are the external conditions that could exacerbate the situation or harm organizations if not addressed. The most obvious threat is the adversary’s progress: the “harvest now, decrypt later” threat is already active – nation-state attackers are currently stealing encrypted data \cite{b9, b1}. This means each passing month that enterprises haven’t upgraded critical encryption is another month of exposure; by the time they do, some data might already sit in an enemy’s vault awaiting decryption. Another looming threat is a potential breakthrough in quantum computing sooner than expected (or even just sooner than organizations are ready for). If, hypothetically, a CRQC capable of breaking RSA-2048 came online by 2030 (on the early side of predictions) \cite{b4}, that could catch many enterprises mid-migration or even not started, leading to security failures and scramble. Even before a full CRQC, criminal exploitation of weakened crypto during transition is a threat – e.g., as hybrid modes roll out, criminals might find clever attacks on poorly implemented hybrids or on systems that temporarily run vulnerable algorithms in parallel. There’s also the threat of regulatory and legal backlash if companies are too slow: once standards are out and government guidance is given, failure to act could later be interpreted as negligence. For example, if a breach happens in 2032 and an enterprise had done nothing about known quantum risks, they might face lawsuits (“you knew this data needed long-term protection and didn’t upgrade”). Another external threat is supply chain risk – an enterprise might upgrade its own systems, but what if its vendors (cloud providers, software suppliers) do not? A chain is as strong as its weakest link; organizations reliant on third-party technology must ensure those suppliers are also transitioning (and this can be hard to guarantee – it’s a threat if key providers lag behind). Additionally, we consider threats to emerging solutions: for instance, if a flaw is discovered in a widely adopted PQC algorithm (similar to the SIKE incident but imagine if it happened to Kyber or Dilithium), that could sow confusion and delay migrations (as people ask “did we bet on the wrong horse?”). Adversaries might even hype such events or use disinformation to reduce trust in new algorithms – an information warfare angle. Finally, quantum computers themselves could introduce new threats beyond cryptography – e.g., quantum algorithms might help attackers in other ways (like solving certain optimization problems that make password cracking or code analysis easier, though the main focus is crypto). The intersection of quantum computing with AI could also amplify threats (AI to analyze decrypted troves or find new vulns). While speculative, these underscore that the threat landscape around quantum is dynamic.
\end{itemize}

The SWOT analysis thus reflects a state where strengths and opportunities provide reason for optimism that this transition can be managed – technology is not the bottleneck – but weaknesses and threats highlight serious concerns that must be addressed promptly. Enterprises have much work to do internally (planning, talent, system upgrades) under the pressure of an external countdown (adversaries moving forward, quantum hardware improving year by year).

\section{Conclusion and Recommendations}
\label{sec:6}
\subsection{Is Enterprise Ready for Quantum-Safe Cybersecurity?}
Given the evidence and analysis, the question “Are enterprises ready for cybersecurity solutions based on quantum computing?” does not have a simple yes or no answer. A more accurate conclusion is a segmented assessment: a small fraction of enterprises – typically large, regulated or highly security-conscious organizations – are taking substantial steps toward quantum readiness, but the vast majority are in the very early stages or have not begun, leaving a significant readiness gap. In 2025, overall enterprise readiness can be characterized as nascent and lagging behind the pace of technological and threat developments.

From the technology perspective, we conclude that readiness of solutions is not the primary bottleneck. Post-quantum cryptography has advanced to the point of practicality: robust standards (e.g., NIST’s PQC algorithms) are in place and reference implementations exist \cite{Yesina2022}. Transitioning to these algorithms is a challenge but a manageable one, akin to past crypto transitions (such as from SHA-1 to SHA-256, or from 1024-bit RSA to 2048-bit RSA) albeit on a larger scale. QKD remains specialized; its adoption will likely be limited to governments and select industries in the near term. So, enterprises do not lack tools – they lack adoption of those tools. In other words, the technology has (or shortly will) reach enterprise-grade maturity, but enterprise utilization is lagging.

From the enterprise readiness perspective, our research indicates that most organizations are not ready if a sudden quantum threat were to materialize. Key indicators like the proportion of companies with a quantum-safe crypto strategy (~5\%) \cite{b1} or those planning budget in the next 1–2 years (~20–40\% among early adopters in best-case sectors \cite{b7}) show that preparedness is more the exception than the norm. Many enterprises are in a state of quantum complacency, consciously or unconsciously betting that they can afford to wait. This is a precarious position, especially for organizations holding long-lived sensitive data (which might already be targeted by adversaries). The enterprise segment that is relatively more ready – e.g., some big banks, telecoms, government contractors – demonstrates that readiness is achievable with top-down support and early planning. These organizations often treat quantum risk as part of their strategic risk register and have begun the arduous multi-year journey of upgrading cryptography. They will likely be the least disrupted when PQC becomes the new norm. In contrast, enterprises that delay may find themselves having to rip-and-replace cryptographic systems under duress, which can be far costlier and riskier.

From the threat actor perspective, we conclude that the urgency is real but not widely internalized by defenders. Adversaries, particularly well-resourced nation-states, are acting on the assumption that quantum advantage in cryptanalysis is coming – their current operations (data theft for future use) bear that out \cite{b4}. Meanwhile, defenders (enterprises) for the most part behave as if this is a problem for tomorrow. This mismatch is dangerous: it means the window of opportunity for defenders to prepare quietly is being wasted. The consensus view that a CRQC is unlikely before 2030s \cite{b5} could lull some into a false sense of security (“we have 10+ years”). But as our analysis highlights, 10 years is likely just enough to get ready if one starts now – not a time to relax. In fact, if we align the estimated earliest appearance of a CRQC (\textasciitilde2030) \cite{b4} with the expected time needed to fully deploy new cryptography across global digital infrastructure (also on the order of a decade or more \cite{b5}), it’s clear we are in a tight race. Enterprises that are not yet moving are essentially banking on the hope that the quantum threat is farther out or that others (governments, vendors) will solve the problem for them in time.

In conclusion, enterprise readiness for quantum cybersecurity in 2025 is inadequate overall, with notable pockets of progress. We can analogize it to climate change preparedness – some forward-looking entities have robust plans and are executing them, but many are doing little more than acknowledging the issue. The likely outcome if this doesn’t change is a scramble: as we approach the late 2020s, more organizations will realize they are behind and will rush to implement PQC (driven either by regulatory mandate or a significant quantum computing milestone announcement). That rush could incur higher costs and potential mistakes, compared to a measured transition starting now.

That said, we should also acknowledge positive momentum: The first movers (in finance, telecom, government) will serve as testbeds, ironing out kinks in PQC deployment and sharing lessons. The vendor ecosystem is rapidly evolving, which will make it easier for latecomers to adopt quantum-safe solutions. We might see a scenario where, by around 2027–2028, quantum-safe options are standard in most products (Microsoft, Amazon, Google, etc., all providing them by default), so enterprises can upgrade by simply applying updates. This will certainly help mid-to-small organizations that lack internal expertise. In that optimistic view, even lagging enterprises could catch up just in time, especially if they have maintained good general cybersecurity hygiene (e.g., ability to roll out patches enterprisewide, etc.). However, critical sectors with long asset lifespans (power grids, military systems, healthcare record systems) do not have the luxury of a quick patch; they need forethought now.

Thus, our nuanced answer is: Enterprises are starting to ready themselves for quantum cybersecurity, but readiness is highly uneven. Leaders in this space are perhaps “quantum aware and beginning to transition,” whereas a large tail of organizations remain “quantum vulnerable” due to inaction. If the question is framed as a binary: today, are enterprises ready? – the answer leans toward “No, not yet” for most. But if asked can they be ready by the time they truly need to be? – the answer can be “Yes, if proactive steps are taken immediately and earnestly.” The following recommendations outline what those steps should entail.

\subsection{Roadmap to Quantum Readiness}
Based on our analysis, we propose a set of actionable recommendations for enterprises to enhance their quantum cybersecurity readiness. These recommendations address the identified weaknesses and threats, leveraging strengths and opportunities to ensure a smoother transition to a quantum-safe security posture:

\begin{itemize}
    \item Develop a Quantum Security Roadmap and Governance Structure: Every organization should create a formal roadmap for transitioning to quantum-safe cryptography. This plan should include timelines (aligned with external guidance like NIST or NCSC milestones), budget estimates, and roles/responsibilities. Senior management and the board should be briefed on quantum risks and endorse the plan, treating it as part of the enterprise risk management strategy. Establish a governance structure – for example, a Crypto-Agility Task Force or assign the responsibility to an existing security committee – to oversee execution of the roadmap. This ensures accountability. Given that only 5\% have a strategy now \cite{b1}, simply formulating one is the first leap for the other 95\%. The roadmap should prioritize systems based on the sensitivity and longevity of data they protect: identify “must fix first” areas (e.g., VPN that protects archive backups, PKI used for code signing, etc. where compromise would be catastrophic). Set target dates (perhaps initial PQC pilots by 2025–26, broader rollout by 2027–28, completion by 2030 in critical areas, aligning with the notion of having mitigations by 2031 as NCSC suggests \cite{b7}).

    \item Conduct a Cryptographic Inventory and Risk Assessment: Organizations need to know what cryptography they have and where. Use automated tools and manual audits to inventory all uses of cryptography in the enterprise – including protocols, applications, data at rest, data in transit, and third-party services. For each, determine which algorithms are used and whether they are quantum-vulnerable (RSA, DH, ECC, DSA, etc.). Also note constraints – e.g., an IoT device with RSA that cannot easily be upgraded. Complement this with a quantum risk assessment: categorize data by shelf-life and sensitivity. Data that needs confidentiality beyond X years (where X might be 5-10 years or more) should be marked as at-risk for quantum exposure. This assessment identifies which data or systems absolutely must be quantum-safe sooner. For example, patient healthcare data (HIPAA requires retention ~6 years) or state secrets (could be sensitive for decades) would top the list. This process will reveal the scope of the problem and help in planning migrations and possible mitigations (like doubling symmetric keys as an interim step). It also inherently forces the organization to confront its crypto-agility (or lack thereof) and start cataloguing where upgrades might be needed.

    \item Embrace Crypto-Agility in Architecture and Procurement: Going forward, design and refactor systems to be crypto-agile – decouple algorithm implementations from business logic, support multiple algorithms, and be able to swap out algorithms via configuration. When procuring new software or hardware, include requirements for crypto-agility and/or PQC support. For example, an RFP for a new VPN solution could mandate compliance with Suite 3 (PQC algorithms) or at least a roadmap to support them. Enterprises should also demand transparency from vendors on their quantum-safe plans – many vendors now have statements or “quantum readiness” roadmaps, and if a critical vendor has no such plan, that’s a red flag. By building crypto-agility now, organizations “future proof” themselves not just against quantum but any future cryptographic changes. One practical step is to implement support for hybrid cryptographic modes in applications (where feasible) and begin testing them internally – for instance, set up a test website or service that uses a hybrid TLS (classical + PQC) and see how it interacts with clients. Early testing can uncover compatibility issues in a low-stakes environment.

    \item Prioritize Early Implementation of PQC in High-Value Areas: Don’t wait for an all-at-once switch; identify a few key areas where you can start using post-quantum algorithms sooner rather than later. Many organizations are doing this for their internal PKI or secure communications. For example, a company could deploy a PQC-capable VPN for backups between data centers (with both classical and PQC running in parallel for safety initially) \cite{Aquina2025}. Or begin using PQC-based code signing for software that will be in use for a long time (to ensure its signatures remain valid into the future). Another candidate is encrypted archives: if you are storing sensitive data encrypted with RSA, consider re-encrypting with a quantum-safe algorithm or a large symmetric key, to hedge against future decryption. These early implementations serve as learning opportunities – they reveal performance impacts, integration challenges, and give the security team hands-on experience. Additionally, starting small mitigates risk; you wouldn’t want to flip your entire customer-facing infrastructure to PQC overnight without experience. Instead, do it in contained environments first. Notably, as NIST’s Dustin Moody said: \textit{“there is no need to wait… Go ahead and start using these [PQC algorithms]}” \cite{b19} – the standards are ready. Ensure however that any early adoption is done with the latest versions and security patches of libraries (since PQC is new, bugs might still be found and fixed).

    \item Invest in Training and Talent Acquisition: Address the talent weakness by building a pipeline of quantum-savvy security expertise. This involves multiple tactics: sponsor current security engineers to obtain training/certifications in cryptography and PQC (courses, workshops, perhaps universities now offering “post-quantum cryptography” programs). Encourage CISOs and architects to follow publications from NIST, ISACA, etc., on quantum readiness to stay current. For bigger organizations, hiring a cryptographer or a quantum risk specialist could be warranted – someone who can lead the crypto modernization effort full-time. If that’s not feasible, consider partnerships or consulting engagements with firms that specialize in this domain to jump-start knowledge transfer. According to industry findings, the window to secure quantum-savvy talent is closing as demand will surge \cite{b7}. So acting early could also be a competitive advantage in hiring – once everyone is scrambling, talent will be scarce and expensive. Also collaborate with industry groups or join pilot projects (some sectors have cross-company quantum readiness projects where you can send delegates to learn and contribute). By developing internal skills, the enterprise not only addresses the current migration but also positions itself to handle future developments (like if new algorithms emerge or new threats like quantum attacks on other systems appear).

    \item Monitor Threat Intelligence and Progress in Quantum Computing: Stay informed about advancements in quantum computing that could alter the risk timeline. This means monitoring reports from bodies like the U.S. National Quantum Coordination Office, academic breakthroughs, and also any classified or proprietary intel if available (some companies in defense or critical infrastructure might get government intel updates). If, say, a breakthrough in error correction is announced that moves the timeline up, the enterprise can then accelerate its plans accordingly. Also track what threat groups are doing – e.g., if intelligence indicates a particular adversary is aggressively harvesting encrypted data from your sector, that should raise priority (perhaps already the case as per Volt Typhoon example \cite{b9}). Some specific actions: subscribe to alerts or newsletters focused on quantum risk (CISA, NCSC often publish guidance; the U.S. NSA has a regularly updated FAQ on quantum-resistant cryptography \cite{b38}). Internally, incorporate quantum scenarios in threat modeling and red-teaming exercises: for instance, ask “what if our VPN traffic were decrypted by an adversary in 5 years – what could they do with that?” to identify which data would be most damaging if exposed.

    \item Collaborate and Share Best Practices: Because quantum-readiness is a broad ecosystem challenge, enterprises should collaborate rather than work in isolation. Join industry consortia or information-sharing groups focusing on quantum security (for example, the Quantum Economic Development Consortium (QED-C) has a cybersecurity working group). Share your lessons learned from any pilot implementations in exchange for others’ insights – if a bank finds a performance issue with a certain PQC algorithm in their transaction system, that info could help another bank avoid the same pitfall, and vice versa. Professional organizations like ISACA, (ISC)², IEEE are building knowledge repositories – contribute to and draw from those. This collaboration can also amplify the voice of enterprises in pushing vendors and standards bodies: a group of large companies together can, for instance, request a cloud provider to prioritize adding PQC support, whereas one company alone might not sway as much.

    \item Consider Quantum-Safe Encryption for Long-Term Data Now: For data that absolutely must remain confidential for many years (think of state secrets, critical intellectual property, or personal data like DNA information that could have lifelong impact), consider adding quantum-safe protection immediately. This might mean using very large symmetric keys (e.g., AES-256 or beyond, since symmetric is relatively safe against quantum aside from Grover’s algorithm) or using combinatorial approaches (encrypt data under multiple schemes – e.g., classical + PQC – and store both, so an attacker would need to break both). Some organizations have started doing “crypto layer-ing” where sensitive backups are encrypted with one of the new NIST PQC KEMs in addition to classical encryption. The cost is low (just some additional computation/storage) and it acts as insurance if quantum capabilities come sooner. Another approach is secret splitting or quantum one-time pads for the most crucial data, if feasible (though key distribution for that can be a challenge, QKD could help in some niche cases, by delivering one-time pad keys). The idea is to reduce the exposure of crown jewels as much as possible in these intermediary years.

    \item Update Incident Response and Key Management Processes: As the environment changes, ensure that security policies and procedures are updated. For example, incident response plans should account for scenarios like “encrypted data was stolen – how do we assess the risk that it could be decrypted later?” Even though one might not feel the impact immediately, it should perhaps be treated as a major incident given future implications. In key management, start phasing in algorithms: update Certificate Authority processes to be able to issue PQC-based certificates; maintain parallel hierarchies if needed. Ensure cryptographic agility is tested – do disaster recovery drills where you simulate discovery of a broken algorithm and have to swap it out (this will test your preparedness for an event like a sudden crack of an algorithm or a sudden quantum breakthrough). Also, as PQC involves different key lengths and formats, verify that backup systems, HSMs (Hardware Security Modules), etc., can handle them; if not, plan upgrades.

    \item Leverage Government and Framework Guidance: Finally, use the available frameworks to guide your journey. For instance, NIST is expected to release a special publication on migration to PQC (a draft NIST SP 800-xxx Transition to PQC is already out) \cite{b27} – follow its recommendations. Governments often publish roadmaps or whitepapers (the UK, U.S., EU, etc.), which contain checklists and timelines. Adopt those as benchmarks. A concrete example: The U.S. CISA in 2023 launched a “PQC Initiative” urging all organizations to get ready and provided a project plan. By aligning with such frameworks, enterprises can ensure they’re meeting best practices and also demonstrate to regulators/auditors that they are following recognized standards.
    
\end{itemize}

In implementing these recommendations, it’s important to avoid a pitfall: do not attempt a Big Bang overhaul. Instead, use an incremental, risk-based approach. The transition to quantum-safe security is a marathon, not a sprint – but it’s a marathon that has already begun and demands steady pacing. Start small, iterate, and expand.

In conclusion, enterprises that take a proactive, structured approach as outlined above can transform the looming quantum threat from a daunting uncertainty into a manageable (even routine) security upgrade over the coming years. The task is undoubtedly complex, touching virtually every aspect of IT, from applications to infrastructure to policy. But it is also an opportunity – to modernize cryptographic infrastructure, to improve security practices, and to assure customers and stakeholders that one’s organization is forward-looking and trustworthy in safeguarding data for the long term. Those who act now will likely thank themselves later; those who do not may find, in a decade’s time, that they have fallen irreparably behind in the face of a new cryptographic reality.

By combining technological preparedness (deploying PQC, etc.), organizational adjustments (planning, training), and vigilance against threats (monitoring and mitigating HNDL and other tactics), enterprises can answer the question of readiness with increasing confidence: Yes, we are getting ready – and we will be ready when the time comes. Each organization’s journey will differ, but the destination – a quantum-resilient security posture – is now a critical objective on the horizon of cybersecurity. Future research and efforts will continue to support this transition, including the development of more efficient PQC algorithms, techniques for easing migrations (like automated code translation for crypto), and maybe even quantum-resistant architectures for things like blockchain and digital identity. Enterprise leaders should keep abreast of these developments and remain adaptable. In the ever-evolving chess match of cybersecurity, the advent of quantum computing is a major new move – but with foresight and collective action, we can counter it and even harness its benefits securely, ensuring that the foundations of digital trust remain intact in the quantum age.


\bibliographystyle{plainnat} 
\input{main.bbl}

\end{document}